# CRITIQUE ON VINDICATION OF PANSPERMIA

Pushkar Ganesh Vaidya
Indian Astrobiology Research Centre (IARC)
pushkar@astrobiology.co.in

**Abstract**

In January 2001, air samples were collected from Earth's stratosphere. From these air samples, cultures of three microorganisms were obtained. It was reasoned that these microorganisms are of cometary origin and thereby cometary panspermia stood vindicated. The fact that these microorganisms had essentially the same characteristics as terrestrial microorganisms was explained using cometary panspermia. Here, the findings are reinterpreted in the light of niche ecology and adaptations. It is asserted that the microorganisms captured from the stratosphere cannot be of cometary origin as they are contrary to the kind of microorganisms one would expect to find in a cometary niche.

## 1. Cometary Panspermia

The hypothesis of cometary panspermia needs a small fraction of microorganisms present in the interstellar cloud from which the solar system formed to have retained viability, or to be capable of revival after being incorporated into newly formed comets. This fraction could be exceedingly small (*N.C. Wickramasinghe et al.* 2003). It has been suggested that radioactive heat sources like 26AL served to maintain a warm liquid interior in the young comets for about a million years (*Hoyle, F. & Wickramasinghe, N.C.* 1985) and this time period was adequate for microorganisms present in the young comets to replicate and occupy a significant volume of a comet.

Some comets owing to orbital disruptions get deflected towards the inner solar system thus carrying microorganisms onto the Earth and other inner planets. Thus, as per cometary panspermia life was first brought to Earth, about 4 billion years ago by comets. Comets have interacted with Earth throughout the past 4 billion years; a strong prediction of the hypothesis of cometary panspermia is that cometary injections of microbial life must be an ongoing process. (*N.C. Wickramasinghe et al.* 2003).

## 2. Balloon Experiment

On 21 January 2001 air samples were collected over Hyderabad using balloonborne cryosamplers. The collection was done in the Earth's stratosphere at various heights up to 41 km (*N. C. Wickramasinghe et al.* 2003). In 2002, from the air samples collected, cultures of three microorganisms were obtained (*Wainwright et al.* 2002) using a soft potato dextrose agar medium;

(a) The coccus (spherical bacterium, often growing in clumps) 99.8% similar (as determined by 16S RNA analysis) to *Staphylococcus pasteuri*

(b) The bacillus (rods), 100% similar (as determined by 16S RNA analysis) to *Bacillus simplex*

(c) A fungus identified as *Engyodontium albus* (Limber) de Hoog (as identified by CABI Bioscience (Egham)

### 2.1 Characteristics of microorganisms isolated from stratospheric samples

It was reported that both the bacteria captured from the stratosphere were grampositive, non-acid fast, catalase-positive, and facultatively anaerobic. The bacterial isolates were reported to exhibit potentially UV-resistant morphologies as the environmental conditions found at 41 km are extreme in terms of UV exposure, low temperatures and pressures.

No organisms were reported to have been isolated using nutrient-free silica gel medium suggesting that oligotrophs were absent or, if present, were incapable of growing under the physical chemical conditions provided by the medium (*Wainwright et al.* 2002).

### 2.2 Extraterrestrial source

It was asserted that the strictest possible anti-contamination measures were taken at every stage of the balloon experiment. So, the microorganisms found in the stratosphere could either be from Earth or from Space. It was stated that only powerful events like volcanoes could propel particles past the tropopause, into the stratosphere; no such event was recorded in the two-year run-up to the balloon launch date on January 20, 2001, and so the microorganisms found in the stratosphere cannot be from Earth. Similar objections held for other rare meteorological events. Therefore, the microorganisms must come from an extraterrestrial source. This extraterrestrial source is comets (*Wainwright et al.* 2002).

## 3. Interpretation of the balloon experiment findings

The findings of the balloon experiment were interpreted as proof for ongoing cometary injections of microbial life to Earth and thereby cometary panspermia stood vindicated. The fact that the microorganisms captured from the stratosphere have essentially the same characteristics as terrestrial microorganisms was explained using cometary panspermia. The 'similarity' of the microorganisms was said to be consistent with panspermia theories in which organisms on Earth are derived from cometary organisms and so the 'similarity' should not be a surprise (*N.C. Wickramasinghe et al*. 2003).

Note - Henceforth, **R** will refer to the 'microorganisms captured from the stratosphere during the balloon experiment, claimed to be of cometary origin and are the $n^{th}$ generation of the microorganisms which replicated in comets during the cometesimal formation period, which lasted for about a million years'.

## 4. Replication of microorganisms in comets

Replication of microorganisms in comets is proposed to have happened during the short period of cometesimal formation, early in the solar system's history, which lasted about a million years and marked by the presence of liquid water, allowing the microorganisms to occupy a significant volume of a comet (*Hoyle, F. & Wickramasinghe, N.C.* 1985). Thus, presence of liquid water is thought to be necessary for replication of microorganisms in comets.

The internal temperatures reached in the cometary nuclei during the period of cometesimal formation, owing to internal heating by radionuclides like 40K, 235U, 238U, 232Th, especially 26AL lie on average between 20 and 120 K (*Meech K. J. & Svoren J*. 2005) and capped at 137K (*Prialnik et al*. 1987).

Brief periods of liquid water availability especially during the short *active phase* may also allow replication. The surface temperature of cometary nuclei can rise by a few hundred-degree Kelvin, reaching the range of 300 to 400 K when they approach the Sun as observed in comets 1P/Halley and 19P/Borrelly (*Keller et al*. 2005). Such temperatures are hot enough to melt water to ice.

Various thermal mechanisms have been suggested by which water may remain in liquid state during the dominant *cold storage phase* (*Podolak & Prialnik,* 2006). Also chance events like collisions can result in brief periods of liquid water availability across all phases.

## 5. Cometary environment

Comets formed in the early stages of the condensation of the solar system. They contain the most pristine material available from that epoch which helps us understand conditions that existed in the young solar nebula more than 4.6 billion years ago. The major comet reservoirs are Oort cloud and Kuiper belt as well as trans-Neptunian scattered disc. Comets spend most of their lives in these reservoirs. The equilibrium surface temperatures in the cometary reservoirs are low ~10 K in the Oort cloud and ~40 K in the Kuiper belt (*Jewitt, D.C*. 2005).

The aging or evolutionary effects that a comet nucleus will experience can be divided into four primary areas: the *precometary phase*, where the interstellar material is altered prior to incorporation into the nucleus; the *accretion phase,* the period of nucleus formation; the *cold storage phase*, where the comet is stored for long periods at large distances from the Sun; and the *active phase* where the comet undergoes drastic changes owing to increased solar insolation as it approaches the inner solar system (*Meech K*.1999) and (*Meech K. J. & Svoren J*. 2005).

Cometary nuclei are a predominantly frozen, dry, dark and anaerobic environments made up of various substances. There is no one fixed composition of comets. The volatile components undergo drastic changes during the *active phase*, consequently affecting the entire physical attributes and chemical composition of the cometary nuclei.

Cometary environment is a microgravity environment, making it unique in comparison to Earth.

**6. Adaptations & Cometary niche**

Species adapt to the ecological niche they occupy. Ecological niche is a term for the position of a species within an ecosystem, describing both the range of conditions necessary for persistence of the species, and its ecological role in the ecosystem (*Polechová & Storch*. 2008).

Extremophiles have adapted to survive in extreme conditions of temperature, acidity, salinity, pressure, toxin concentration etc. The purpose of this paper is not to list classes of extremophiles vis-à-vis their adaptations and reiterate how specific adaptations are required for different extreme ecological niches. (*Please refer the works in the suggested reading section*). The purpose is simply to stress the fact that microorganisms can survive in extreme ecological niches provided they 'adapt' to those niches.

For instance, a microorganism found in extreme cold environments, must show adaptations as found in a psychrophile and not a thermophile. To clarify further, a microorganism occupying an extreme ecological niche must show relevant adaptations although it might be found in environments where the same adaptations are not essential.

Cometary environment is an extreme environment (*refer section 4*) and thus is an extreme niche for any microorganism to occupy. The conditions are so extreme that if a microorganism does not adapt then it will die.

**7. Replication of microorganisms in the Cometary niche**

Replication of microorganisms during the cometesimal formation period (*refer section 4*) which lasted for about a million years, calls upon evolutionary adaptations in order to survive in the cometary niche. The major factors which would drive adaptive evolution are given here.

<u>Microgravity</u>

Gravitational acceleration has been constant throughout the ~4B years of biological evolution on Earth and life forms have evolved to function under a 1-G force. When gravity is altered, biological changes are observed even when cells are isolated from the whole organism and grown in culture. It has been shown beyond doubt that microgravity brings about alterations in cellular function, gene expression and structure (*Morey-Holton, E.R.* 2003).

Example - To study the effects of space flight on microorganisms, tubes containing *Salmonella* were sent as an experimental payload aboard the Space Shuttle Atlantis. Amazingly, compared to bacteria that remained on Earth for a synchronous control experiment, the space-traveling *Salmonella* had changed expression of 167 genes and showed biofilm formation (*Wilson et al.* 1997).

**R** do not show any adaptations towards microgravity (*refer 2.1*).

Cold Temperature

Psychrophiles are extremophiles capable of growth and reproduction at extreme low temperatures. Lad studies have found them to reproduce at 272 K, just below the freezing point of water (*Reid et al.* 2006). Several changes take place inside the cell when it is exposed to low temperature. These include 1) loss of membrane flexibility, 2) stabilization of secondary structures in nucleic acids, 3) increase in negative supercoiling of DNA, 4) unfolding or improper folding and methylation of some proteins and 5) Loss of membrane flexibility affects the membrane-associated functions such as transport (*Phadtare & Inouye.* 2008).

Psychrophiles are too small to insulate themselves from these effects of cold and therefore, the only recourse is to adapt by altering their cellular composition and functions. Adaptations include lipid cell membranes which are chemically resistant to cold induced stiffening, production of range of 'antifreeze' proteins to keep cellular interior liquid and enzymes protected by cryoprotectants (*Russell.* 2008).

Example - *Colwellia psychrerythraea* 34H has been isolated from Arctic marine sediments. It can grow even at temperatures as low as 272 K. It produces extracellular polysaccharides and, in particular, cold-active enzymes with low temperature optima for activity and marked heat instability (*Methe.* 2005).

**R** do not show any adaptations towards low temperatures (*refer 2.1*).

Low Nutrition

Oligotrophs are organisms that live in low-nutrient habitats. Oligotrophic organisms become small as a matter of optimizing their surface-to-volume ratio; this is to enhance nutrient uptake. Low-nutrient habitats make the oligotrophs grow slowly which is manifested as a marked decrease in the production of ribosomes and enzymes (*Van Etten.* 1999).

Example - *S. alaskensis* has been isolated from Alaskan waters; an oligotrophic environment. It shows unique genetic and physiological properties which are fundamentally different from those of the well studied bacteria such as *Escherichia coli* (*Copeland et al.* 2006).

**R** do not show any adaptations towards oligotrophic conditions (*refer 2.1*).

Genetic Make-up

Evolution experiments explicitly show how the genomes and phenotypic properties evolve over generations in microorganism in response to changes in environment (*Elena & Lenski.* 2003). Adaptive genetic differences in relation to the ecological niche of a species can be identified by determining the "selective signature" of a gene; that is, the pattern of fast or slow evolution of that gene across a group of species, and use that signature to map changes to shifts in an organism's environment (*Shapiro & Alm.* 2008).

Example – The genome of *Idiomarina loihiensis*, a marine bacterium adapted to life near sulfurous hydrothermal vents, show an integrated mechanism of metabolic adaptation to the constantly changing deep-sea hydrothermal ecosystem. The genes involved in sugar fermentation, for carbon and energy underwent significant changes over millions of years to help the bacterium obtain carbon and energy through amino acid catabolism ;a necessity for life in that ecological niche. The same genes all completely lost in the psychrophile *Colwellia psychrerythraea* as sugar metabolism is not required (*Shapiro & Alm.* 2008).

**R** do not show any distinct genetic adaptations necessary for a cometary niche (*refer 2.1*).

<u>Energy Source</u>

Energy source refers to the pathways used by an organism to produce ATP, which is required for metabolism. In the absence of light as an energy source a microorganism must explore alternate pathways; chemotrophy is a pathway by which microorganisms obtain energy by the oxidation of electron donating molecules in their environments. These molecules can be organic (organotrophs) or inorganic (lithotrophs) (*Pierson*. 2001).

Example - *Thiobacillus* species obtain energy by "burning" compounds such as $H_2S$, $S$ and $S_2O_3^{2-}$ with oxygen, producing more oxidized forms of sulfur (*Harding & Holbert*. 2000).

**R** do not show any pathways necessary to obtain energy in a cometary niche (*refer 2.1*).

**8. Conclusions**

The three microorganisms captured during the balloon experiment do not exhibit any distinct adaptations expected to be seen in microorganisms occupying a cometary niche.

Therefore, the findings of the balloon experiment are re-interpreted whereby it is asserted that these microorganisms are from a terrestrial source and there exists a hitherto unknown atmospheric phenomenon to account for their presence.

Consequently, cometary panspermia is not vindicated.

**Suggested reading**

Extremophiles: Microbial Life in Extreme Environments by Koki Horikoshi
Wiley-Liss Press (**1998**) **ISBN-13: 978-0471026181**.

Extremophiles, 35 by Fred Rainey and Aharon Oren
Academic Press (**2006**) **ISBN-13: 978-0-12-521536-7**.

Psychrophiles: from Biodiversity to Biotechnolgy by Rosa Margesin
Springer Press (**2007**) **ISBN-13: 978-3540743347.**

**Acknowledgments**

Dr. Arun Dholakia, Surat and Sunil Churi, Mumbai for their support.
Hemal, Sandeep and Vikram, Mumbai, for scientific discussions.
Dr. A.P.J. Abdul Kalam for the moral support.